\def\Obh2{\Omega_{B} {h}^2}

\documentstyle[epsfig,prd,aps,twocolumn]{revtex}
\begin{document}
\draft
\twocolumn[\hsize\textwidth\columnwidth\hsize\csname
@twocolumnfalse\endcsname

\pagestyle{empty}
%\begin{center}
\bigskip

\rightline{FERMILAB--Pub--00--239-A}

\title{\Large \bf What Is The BBN Prediction for the Baryon Density\\
and How Reliable Is It?}
\bigskip

\author{Scott Burles,$^{1,2}$ Kenneth M. Nollett,$^{4,5}$ and
Michael S. Turner$^{2,3,4}$}

\address{
{$^1$Experimental Astrophysics Group, Fermi National Accelerator Laboratory,\\
Box 500, Batavia, IL~~60510-0500}\\
{$^2$Department of Astronomy \& Astrophysics \\
Enrico Fermi Institute, The University of Chicago, Chicago, IL~~60637-1433}\\
{ $^3$NASA/Fermilab Astrophysics Center\\
Fermi National Accelerator Laboratory, Batavia, IL~~60510-0500}\\
{$^4$Department of Physics\\
Enrico Fermi Institute, The University of Chicago, Chicago, IL~~60637-1433}\\
{ $^5$Physics Division, Argonne National Laboratory, Argonne, IL 60439}}
\date{\today. Submitted to Phys. Rev. D.}

\maketitle

\begin{abstract}
Together, the standard theory of big-bang nucleosynthesis (BBN) and
the primeval deuterium abundance now very precisely peg the baryon
density.  Based upon our analysis of the deuterium data and the
theoretical uncertainties associated with the BBN predictions, we
determine $\Omega_B h^2 = 0.020\pm 0.002$ (95\% C.L.), with the
uncertainty from the measured deuterium abundance about twice that
from the predicted abundance.  We discuss critically the reliability
of the BBN baryon density, and in light of {\em possible}
systematic uncertainties also derive a very conservative range.  
We conclude that within the standard cosmology and standard theory of
BBN a baryon density $\Omega_Bh^2=0.032$ (the central value implied by
recent CMB anisotropy measurements) simply cannot be accommodated.
\end{abstract}

\pacs{26.35.+c, 98.80.Ft, 98.80.Es}

]

%\newpage
%\pagestyle{plain}
%\setcounter{page}{1}
%\newpage

\section{Introduction}  Since the determination of its solar
system abundance by the Apollo astronauts almost thirty years ago,
deuterium has been used to constrain the density of ordinary matter
\cite{apollo,yorketal}.  The reason is simple: big-bang deuterium
production has a strong dependence upon the baryon density ($\propto
\rho_B^{-1.6}$), and astrophysical processes since have only destroyed
deuterium \cite{dns}.

Until recently, deuterium was used to set an upper limit to the baryon
density (around 10\% of critical density), based upon the fact that
the big bang must produce {\em at least} the amount of deuterium seen
in the local interstellar medium (ISM).  Together with measurements of
the abundances of the other light elements produced in the big bang, a
concordance interval for the baryon density was derived, $0.007 \le
\Omega_Bh^2 \le 0.024$ \cite{copietal}.  For two decades the big-bang 
nucleosynthesis (BBN) baryon density has stood as the best
determination of the amount of ordinary matter and the linchpin in the
case for nonbaryonic dark matter \cite{mst_review}.

A dramatic change occurred in 1998 when the abundance of deuterium was
measured in high-redshift clouds of pristine gas backlit by even more
distant quasars.  Tytler and his collaborators have now determined
deuterium abundances for three clouds and derived upper limits for a
number of other clouds
\cite{tytler2000,omeara}.  Based upon these and other results we infer a
primeval deuterium abundance,
(D/H)$_P=(3.0\pm 0.4)\times 10^{-5}$ (95\% C.L.), which
leads to a precision determination of the baryon density \cite{bnt}:
\begin{equation}
\Omega_B h^2 = 0.020\pm 0.002 \ \ (95\%\ {\rm C.L.}).
\end{equation}
$\Omega_B$ is the fraction of critical density contributed
by baryons, $h=H_0/100\,{\rm km\,sec^{-1}\,Mpc^{-1}}$, and the
physical baryon density $\rho_B = 1.878(\Omega_Bh^2) \times 10^{-29}\,
{\rm g\,cm^{-3}}$. 

A key test of the BBN prediction, and indeed the consistency of the
standard cosmology itself, lies ahead.  Measurements of cosmic
microwave background (CMB) anisotropy on small angular scales can
ultimately determine the baryon density to an accuracy of around 1\%.
The physics involved is very different: gravity-driven acoustic
oscillations of the photon-baryon fluid when the Universe was around
500\,000 years old.  The first step toward this important goal was
taken recently when the BOOMERANG and MAXIMA CMB experiments reported
results for the baryon density: $\Omega_B h^2 =
0.032^{+0.009}_{-0.008}$ (95\% C.L.)
\cite{boom_max}.  At about the 2$\sigma$ level, this independent measure
of the amount of ordinary matter agrees with the BBN prediction and
supports the longstanding BBN argument for non-baryonic dark matter.

The CMB determination of the baryon density should improve
dramatically over the next few years, making a very precise comparison
of the two methods possible.  The difference between the BBN
and CMB baryon densities has already triggered lively discussion in
the literature \cite{discussion}.  Motivated by this, we have written
this paper to explain the BBN prediction for the baryon density and
the associated uncertainties.

\begin{figure}
\centerline{\epsfig{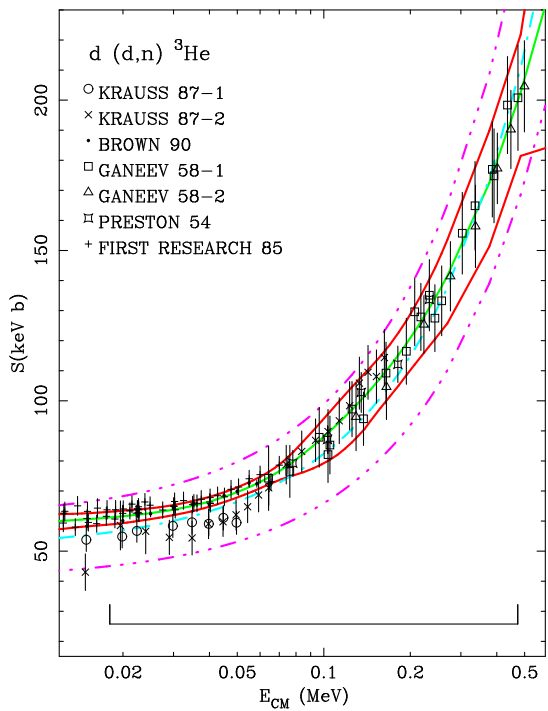}}
\caption{Cross-section data and fits for one of the processes
important for determining the BBN deuterium yield.
Solid curves indicate best-fit and 95\% C.L. limits from
our Monte Carlo method \protect\cite{bntt,nb}; broken curves
indicate the corresponding best-fit and 95\% C.L. curves from
earlier work \protect\cite{SKM}.  The bracket at the bottom
indicates the energy range where this cross section is
needed in order to compute all light-element abundances to
an accuracy of one-tenth of their current uncertainties.}
\label{fig:ddn}
\end{figure}

\section{Uncertainties in the predicted deuterium abundance}
The standard scenario for big-bang nucleosynthesis begins with the
assumptions of (i) the isotropic and homogeneous
Friedmann-Robertson-Walker (FRW) cosmology, (ii) three massless (or
very light) neutrino species, (iii) zero (or very small) chemical
potentials for the neutrino species, and (iv) spatial homogeneity of
the baryon density.  Within the standard theory, the predictions for
the light-element abundances depend only upon the baryon density (more
precisely, baryon-to-photon ratio) and the dozen nuclear-reaction
rates that enter the calculation.

Essentially all of these rates have been measured at energies relevant
for BBN, and because large numbers of measurements exist for most of
the cross sections, reliable estimates of the uncertainties can be
made (see Fig.~\ref{fig:ddn}).  We have used {\em all} the extant
nuclear data to estimate directly the uncertainties in the BBN
predictions \cite{bntt,nb}.  We integrated the nuclear data over
thermal distributions to obtain thermally averaged rates.  Then,
uncertainties in the predictions were estimated by a Monte Carlo
method, repeatedly calculating the abundances with individual
cross-section data being varied according to their quoted experimental
uncertainties (also accounting for correlated normalization errors).
From the resulting distribution of predicted abundances we have
derived the uncertainties.

\begin{figure}
\centerline{\epsfig{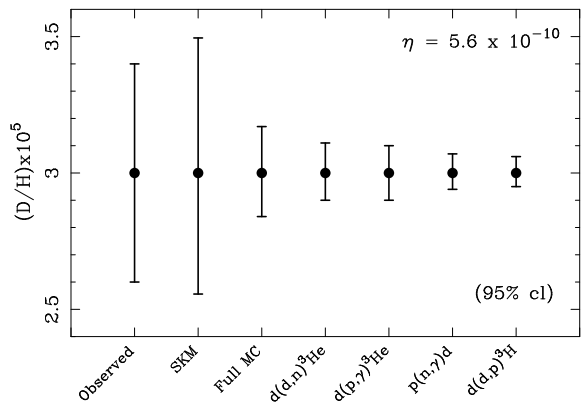}}
\caption{D/H uncertainties (from left to right):
measured D/H \protect\cite{tytler2000}, uncertainty estimated by SKM
\protect\cite{SKM}, our Monte Carlo uncertainty,
and uncertainties arising from the individual reactions, all at
95\% C.L. and for $\Omega_Bh^2 = 0.020$.}
\label{fig:errorbars}
\end{figure}

The uncertainty in the D/H prediction around $\Omega_B h^2 = 
0.020 $ arises mainly from the reactions $d(d,n)^3{\rm He}$,
$d(p,\gamma)^3{\rm He}$, and $p(n,\gamma)d$, with a smaller
contribution from $d(d,p)t$ (see Fig.~\ref{fig:errorbars}).  The total
uncertainty from all of these combined is $5.4\%$ at 95\% C.L., and
increases slightly for larger $\Omega_B h^2$.  Combining this with the
slope of the power-law dependence of D/H on baryon density, we obtain
a nuclear contribution to the uncertainty in the deuterium-derived
baryon density of 3.4\% (95\% C.L.).

We end this discussion with two comments.  First, BBN actually
determines the baryon-to-photon ratio when the Universe was about 100
seconds old, $\eta = (5.6 \pm 0.6)\times 10^{-10}$ (95\% C.L.);
to translate this
into a baryon density today two things are needed: (i) average mass
per baryon; and (ii) the assumption of adiabatic expansion since BBN
\cite{manoj}.  At the 1\% level, the average mass per baryon depends
upon chemical composition;  for the post-BBN primordial mix
or universal solar abundance, $\bar m = 1.670-1.671\times
10^{ -24}\,$g.  Adiabatic expansion since BBN is a feature of the 
standard cosmology.  With these 
assumptions, $\Omega_Bh^2 = (3.650\pm 0.008)\times 10^7\eta$ (95\% C.L.),
where the error comes from the uncertainty in the CMB temperature,
$T=(2.725\pm 0.002)\,$K (95\% C.L.) \cite{mather}.

The second comment is an explanation of why our estimates of the light-element
uncertainties are about a factor of two to three smaller than the
previous very thorough study by Smith et al. (SKM) \cite{SKM} in 1993.
Some of it is simply improved measurements.  However, the bulk of the
difference involves technique.  As described, our
analysis used all the data directly, each data point weighted by its
error bar.  SKM estimated $2\sigma$ limits to cross sections by
constructing envelopes (motivated by theory) forced to contain most of
the measurements, and then from these, derived overall rate
uncertainties.  Although much simpler to implement, this technique
gives too much weight to experiments with large error bars and to
cross-section data in energy intervals that are not as important to
BBN, and results in unnecessarily conservative error estimates.

\section{Uncertainties in the observed deuterium abundance} 
\begin{figure}
\centerline{\epsfig{file=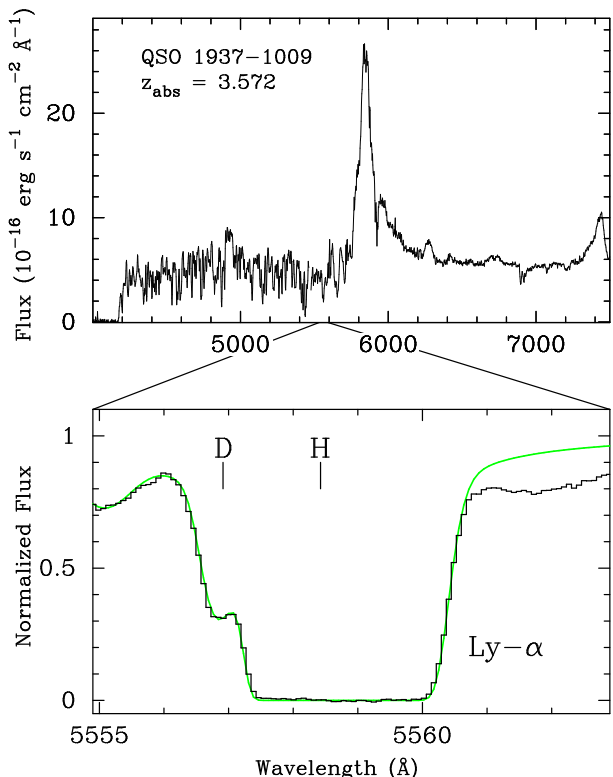,width=8.6cm,angle=0}}
\caption{Spectrum of Q1937-1009; blueward of the
characteristic Lyman-$\alpha$ emission line of the quasar
is the ``forest'' of Lyman-$\alpha$ absorption due to the hundreds
of intervening gas clouds.  The lower panel shows a blowup of the
region around the deuterium detection, a cloud at redshift $z=3.572$,
and the model fit.
}
\label{fig:1937}
\end{figure}

In 1976 Adams
\cite{adams} pointed out that the primordial abundance of deuterium
could be determined by observations of Lyman-series absorption of
quasar light by intervening high-redshift ($z > 2.5$) clouds of
pristine gas, by using the fact that the deuterium feature is
isotopically shifted by 82\,km/s to the blue (see
Fig.~\ref{fig:1937}).  Successfully implementing this idea awaited the
advent of the 10-meter Keck telescope and its HIRES spectrograph
\cite{vogt94}.  At present, four absorbing gas clouds place stringent
constraints on the primeval deuterium abundance, and another five
provide independent consistency checks.  Here we briefly summarize the
state of the observations, and discuss possible systematics.  For a
detailed review, see Ref.~\cite{tytler2000}.

The technique is simple and straightforward; the relative abundances
of deuterium and hydrogen are measured through absorption profile
fitting of the Lyman series.  In these clouds of virtually primordial
material (metal abundance less than 1\% of solar), the relative column
densities of neutral deuterium and neutral hydrogen yield the primeval
D/H ratio without correction for ionization or destruction of deuterium
by stars
\cite{jedamzik1997}.
The major obstacle is the discovery of systems
that are suitable for deuterium detection.  To observe the weak
deuterium feature, the neutral hydrogen column density must exceed
about $10^{17}\,{\rm cm^2}$, and such clouds are relatively rare.  The
current rate of success is approximately 4 suitable clouds per 100
quasars studied.  (Note, for each high-redshift quasar there are hundreds
of intervening gas clouds, but typically only one with sufficiently
high column density to see deuterium.)

There are three systems at $z>2.5$ where deuterium has been detected:
D/H = $(3.3 \pm 0.6)\times 10^{-5}$ \cite{bt98a}; $(4.0 \pm 1.4)\times
10^{-5}$ \cite{bt98b} and $2.5 \pm 0.5$ \cite{omeara} (95\% C.L.).
The third and newest system has a much higher hydrogen column density,
and deuterium is seen in Lyman-$\beta$, $\gamma$, 5, 6, and 7.  A
fourth system provides a strong upper limit: D/H $< 6.7\times 10^{-5}$
(95\% C.L.) \cite{kirkmanetal}.

Using the likelihood distributions for the first two detections and a
gaussian likelihood for the new detection, we infer (D/H)$_P = (3.0
\pm 0.4) \times 10^{-5}$ (95\% C.L.).  We note that the dispersion of
the three detections is somewhat larger than one might expect based
upon the estimated error of the mean, though not unreasonably so: The
reduced $\chi^2 = 7.1$ for two degrees of freedom has a probability of
3\%.

\begin{figure}
\centerline{\epsfig{file=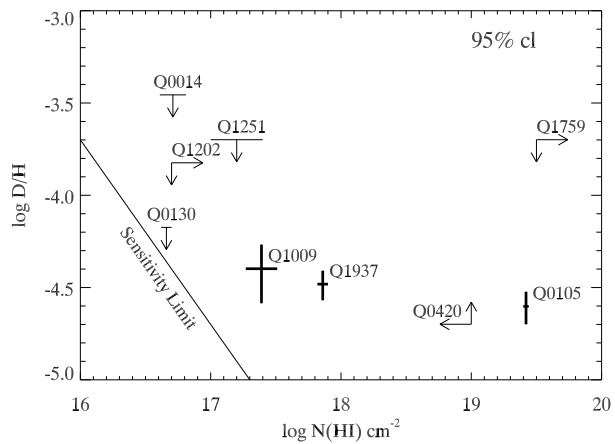,height=6.0cm,angle=0}}
\caption{Summary of deuterium detections and upper limits as
a function of the hydrogen column density (all at 95\% C.L.).
The diagonal line indicates the minimum deuterium abundance that
could be detected as a function of hydrogen column density.  Note
that the newest system lies well above the detection threshold.
}
\label{fig:dhsum}
\end{figure}

Another five extragalactic absorbers that give limits on D/H at high
redshift have been analyzed \cite{dhlimits,bt98c}.  They add no weight
to the D/H determination, but are all consistent with (D/H)$_P = 3
\times 10^{-5}$.   Fig.~\ref{fig:dhsum} summarizes the deuterium
detections and upper limits.

Earlier claims of a factor of ten higher deuterium abundance in
an absorption cloud associated with Q0014+8118 \cite{highdh}
were shown to be a result
of misidentification of the putative deuterium feature \cite{burlesetal}.
The aforementioned upper-limit systems also argue strongly against
high deuterium.  In addition, recent analyses of a low redshift system
suggestive of high D/H \cite{only-lya,tytler1999} are unconvincing due to
lack of spectral coverage of the entire Lyman series.

Although the technique is simple and direct, there are two important
sources of systematic uncertainty.  The first arises from ``hydrogen
interlopers,'' low column-density gas clouds that are coincidentally
situated to mimic deuterium absorption
\cite{bt98c,tytler1999,jedamzik1997,steigman1994}.  Interlopers result
in a one-sided systematic bias which can lead to an overestimation of
D/H.  Absorbing clouds are ubiquitous at high redshifts \cite{lya} and
the chance probability of an interloper is non-negligible.

The probability of an interloper depends on many factors, including
redshift, neutral hydrogen column density, and velocity dispersion as
well as the intrinsic value of D/H.
Here, we present an {\em a posteriori} estimate of the probability
for hydrogen contamination at the level of 10\% or more in each system.
In the three absorption systems with detections, the {\em a priori}
probability for 10\% contamination is: $0.007$, $0.01$ and $0.001$
respectively, assuming that the interloper must fall within 5\,km/s
of the expected position of deuterium absorption.  (Because of its
very high hydrogen column density, the third system is particularly
immune to the possibility of an interloper.) Multiplying
these {\em a priori} probabilities by the number of clouds searched
(about 25), we arrive at estimates for the {\em a posteriori} probabilities,
$0.18$, $0.25$, $0.03$ respectively.  The final joint probability
that all three systems are contaminated by more than 10\%
is less than 0.1\%.  

This small probability, and the consistency of the three
deuterium detections, argues strongly for deuterium detections rather
than hydrogen interlopers.  One can attempt to correct for hydrogen
contamination by assuming a flat prior for contamination of 0 to 10\%
in the lower column-density systems, Q1009 and Q1937.  The likelihood
for (D/H)$_P$, in this case, has a central value that is 3\% lower and
an uncertainty that is slightly (20\%) larger.  This analysis further
suggests that contamination by interlopers is not a significant
systematic; further, allowing for this small contamination lessens the
possible significance of the larger than expected scatter in the
derived deuterium abundances.

The other important source of systematic error involves model fitting
of absorption line profiles in quasar spectra.  A small number of Voigt
profiles were used to model the deuterium absorption systems
\cite{bt98a,bt98b}.  The effect of model-parameter choice has been
studied extensively in the measurement of D/H at high redshift
\cite{bt98c,burlesthesis}, as well as in the local interstellar medium
\cite{dh_ism,vidal-madjar}.  Including the uncertainty of the
unabsorbed quasar continuum, the estimated possible systematic bias is at
the level of 5\%.  This bias could affect all sightlines, but it
should have a different magnitude and direction from one system to the
next.

Finally, it has been argued that astrophysical mechanisms that
significantly alter the deuterium abundance in these clouds are
implausible \cite{jedamzik1997}; the relatively small intrinsic
scatter allowed by the current data argues against any significant
post-BBN production or destruction, as more scatter would be expected.
This argument is weaker than those above, since there could be
some real scatter in the data.

\section{Why BBN Cannot Accommodate $\Obh2 = 0.03$}
Because the uncertainty associated with the CMB baryon density is
significant (about 15\%), there is nothing special about the central
value, $\Omega_Bh^2 = 0.032$.  Nonetheless, it is interesting and
instructive to ask if BBN could accommodate such a high value.  The
answer is a resounding no for three reasons: D, $^7$Li and $^4$He, in
that order.  For a baryon density of $\Omega_Bh^2 = 0.03$, the
predicted abundances are: (D/H)$=(1.6\pm 0.12)\times 10^{-5}$;
($^7$Li/H) $=(8.5\pm 0.7)\times 10^{-10}$; and $Y_P=0.251\pm 0.001$
(95\% C.L.).  All three conflict significantly with observations.

Aside from high-redshift hydrogen clouds, the most precise
determination of the deuterium abundance is that measured with the HST
for the local ISM, (D/H)$=(1.5\pm 0.2)\times 10^{-5}$ (95\% C.L.)
\cite{dh_ism}.  The technique is the same, except on a smaller scale:
absorption by clouds of neutral gas along the lines of sight to many
nearby stars is measured.  (Variations in the local D/H, both upward
and downward, have been reported and may be statistically significant
\cite{vidal-madjar,local_variation}.  If real, they probably reflect
the inhomogeneity of the local ISM.)  The ISM value for D/H is
essentially equal to the predicted primordial abundance for
$\Omega_Bh^2 = 0.03$ and would imply that the local ISM is pristine
material, because astrophysical processes since BBN have only
destroyed deuterium.  This is in stark contrast to the abundance of
heavy elements (around 2\%) which indicates about half the material
has been processed through stars, and thus implies a primeval
deuterium abundance of about $3\times 10^{-5}$.  Further, the D/H
prediction at $\Omega_Bh^2 = 0.03$ is also lower than the abundance
inferred for the pre-solar nebula, (D/H)$=(2.1\pm 1.0)\times 10^{-5}$
(95\% C.L.) \cite{pre_solar}.

The predicted $^7$Li abundance is about six times that measured in the
atmospheres of old pop II halo stars, ($^7$Li/H)$=[1.7\pm 0.1\,({\rm
95\% C.L.}) \pm 0.2\, ({\rm syst})]\times 10^{-10}$ \cite{li_7}.  While
there is lively debate about the possible depletion of $^7$Li in these
stars (by rotationally driven mixing and convective burning), there is
a consensus that any such depletion must be less than about a factor
of two \cite{li_depletion}.  Thus, $^7$Li too is a serious problem for
this baryon density.

Finally, though there is no consensus about the primeval
abundance of $^4$He, and discussion continues about possible
systematic error, $Y_P = 0.250$ is inconsistent with the two
largest compilations of $^4$He measurements.  The primordial $^4$He
abundances inferred from these two studies of
metal-poor, extragalactic H{\sc ii} regions are:
$Y_P = 0.244\pm 0.004$ \cite{izotov} and $Y_P = 0.234\pm 0.004$ 
\cite{oss}, both at 95\% C.L.

\section{A very conservative range for the BBN baryon density}
We believe our stated range for the BBN baryon density, $\Omega_Bh^2 =
0.020\pm 0.002$ (95\% C.L.) is well justified.  However,
two recent reviews have quoted broader 95\% C.L. intervals:
$0.015\le \Omega_Bh^2 \le 0.023$ \cite{dns_vol}
and $0.004\le \Omega_Bh^2 \le 0.021$ \cite{pdg}.
The differences are simple to explain:  The first review
\cite{dns_vol} used the older SKM analysis for the error
in the predicted deuterium abundance (see Fig.~\ref{fig:errorbars}).
The second review \cite{pdg} still allows for the possibility
that the primeval D/H is as large as $3\times 10^{-4}$.
As described above, we believe the case for
high primeval D/H is simply no longer tenable.

The question remains, just how high can the baryon density be pushed.
To answer that question and obtain a ``very conservative range''
for the baryon density, we revert to the older, more conservative SKM
analysis of the input nuclear data -- though unlikely, our Monte Carlo
analysis could be driven by an experiment with understated errors --
and use a very conservative range for the primeval deuterium
abundance, (D/H)$_P = (2.1-3.9)\times 10^{-5}$.

This range for the primeval deuterium abundance is derived from a
weighted average of the three detections, with the error of the mean
being estimated from the standard deviation of the three detections
using the standard formula.  By so doing, we are in effect
disregarding the estimated errors for the individual measurements, and
instead using the dispersion of the measured values to estimate the
error of the mean.  If the statistical errors are larger than
estimated or if there are unknown systematic errors, this approach
might more accurately reflect the underlying errors.  (We note that
O'Meara et al. \cite{omeara}, motivated by the larger than expected
dispersion of the three detections, have advocated a similar approach
for estimating the uncertainty in the primeval deuterium abundance.)

Adding the theoretical and observational uncertainties discussed above
in quadrature, we obtain a very conservative range for the BBN baryon
density: $\Omega_Bh^2 = 0.016 - 0.026$, which is about twice the width
of what we believe to be a well justified 95\% confidence interval.
Precisely because this very conservative range is predicated upon the
possibility of systematic error, or error that is not well quantified
by a Gaussian distribution function, we have not assigned a confidence
level to it.  We also note that at the upper limit of this extreme
range, $\Omega_Bh^2 = 0.026$, all the light-element abundances,
(D/H)$_P= 2.0\times 10^{-5}$, ($^7$Li/H)$=6.6\times 10^{-10}$ and $Y_P
= 0.249$, are uncomfortably different from their inferred values.

\section{Concluding remarks}
A flood of high-precision data -- from CMB measurements to very large
redshift surveys to precision D/H determinations  -- is transforming cosmology.
Soon, no one will remember a time when the phrase precision cosmology was
an oxymoron.  The new data are testing our most
promising ideas about the early Universe as well as the consistency of the
big-bang framework itself.  A comparison between the
BBN and CMB baryon densities will be one of the most important
consistency tests.

The use of BBN to determine the baryon density
is a mature subject.  The predictions have been scrutinized and
the required nuclear data are measured at the relevant energies.
The determination of the deuterium abundance in nearly pristine high-redshift
gas clouds now involves nine systems, and considerable attention has been paid
to systematic error.  While our stated uncertainty in the
BBN baryon density is small, $\Omega_Bh^2 = 0.020\pm 0.002$ (95\% C.L.), it
is well justified (our very conservative range is only twice as
broad).  As more deuterium systems are discovered and analyzed,
the uncertainty will shrink.  Even without improvement in the nuclear data,
a precision of about 4\% (at 95\% C.L.) is possible.

On the other hand, CMB anisotropy measurements have just recently achieved
sufficient angular resolution to probe the baryon density.  The first
result, a 15\% determination of the baryon density,
$\Omega_Bh^2 = 0.032^{+0.009}_{-0.008}$, is encouraging.  It
supports the longstanding BBN argument for nonbaryonic dark matter
and agrees with the BBN baryon density at about the $2\sigma$ level.
It is important to note that the CMB results for $\Omega_Bh^2$
depend upon the number of free cosmological parameters used in the
analysis and the priors assumed for them \cite{discussion}.
The CMB determination will improve significantly over the next few
years as more experiments probe the sub-degree angular scales,
making a more robust comparison with the BBN prediction possible.

As we have emphasized, there are three good reasons why BBN cannot
tolerate $\Omega_Bh^2 = 0.03$: the predicted abundances of D, $^7$Li
and $^4$He; all would conflict significantly with observed abundances.
The standard cosmology and standard BBN cannot -- and probably will
not have to -- accommodate a baryon density this large.

{\em Should} future CMB measurements zero in on a baryon density
higher than our extreme upper limit, $\Omega_Bh^2 = 0.026$,
they would, in our opinion, implicate
nonstandard cosmology or BBN.  Among the possibilities
are:  large neutrino chemical potentials \cite{chem_pot},
entropy reduction since BBN due to exotic physics \cite{manoj},
a decaying tau neutrino \cite{tau_decay}, neutrino
oscillations \cite{bari},
or the inconsistency of the standard cosmology.  At the moment, none
of these possibilities are particularly compelling,
and we will wait to see if they are necessary.

\acknowledgments  This work was supported by
the DOE (at Chicago and Fermilab) and by the NASA (at Fermilab
by grant NAG 5-7092).  We thank Wayne Hu, Gary Steigman,
Max Tegmark, and David Tytler for useful comments.

\end{document}